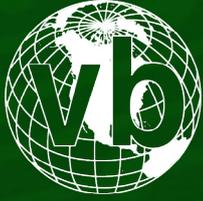

# 2025 BERLIN

# THE ATTRIBUTION STORY OF WHISPERGATE: AN ACADEMIC PERSPECTIVE

Dr. Alexander (Oleksandr) Adamov
*NioGuard Security Lab, Ukraine; Blekinge Institute of Technology (BTH), Sweden; Kharkiv National University of Radio Electronics (NURE), Ukraine*

Dr. Anders Carlsson
*Blekinge Institute of Technology (BTH), Sweden*

ada@nioguard.com
anders.carlsson@bth.se

**ABSTRACT**

This paper explores the challenges of cyber attack attribution, specifically APTs, applying the case study approach for the WhisperGate cyber operation of January 2022 executed by the Russian military intelligence service (GRU) and targeting Ukrainian government entities. The study provides a detailed review of the threat actor identifiers and taxonomies used by leading cybersecurity vendors, focusing on the evolving attribution from *Microsoft*, *ESET* and *CrowdStrike* researchers. Once the attribution to Ember Bear (GRU Unit 29155) is established through technical and intelligence reports, we use both traditional machine-learning (ML) classifiers and a large language model (*ChatGPT*) to analyse the indicators of compromise (IoCs), tactics and techniques to statistically and semantically attribute the WhisperGate attack. Our findings reveal overlapping indicators with the Sandworm group (GRU Unit 74455) but also strong evidence pointing to Ember Bear, especially when the large language model (LLM) is fine-tuned or contextually augmented with additional intelligence. Thus, we show how AI/GenAI with proper fine-tuning is capable of solving the attribution challenge.

**STORYLINE**

On 5 September 2024, Maryland's Grand Jury unsealed a superseding indictment [1] against five members of Russia's Main Intelligence Directorate of the General Staff (GRU) Unit 29155 (a.k.a. Ember Bear) and one previously charged (June 2024) Russian civilian, Amin Stigal [2], for a cyber operation called WhisperGate [3] against the Ukrainian Government services and its allies in the US and Europe – at least 26 NATO countries – on 13 January 2022.

GRU Unit 29155 is associated with APT groups under different names such as Ember Bear (*CrowdStrike*), Saint Bear (*ThreatBook*), Cadet Blizzard (*Microsoft*), FROZENVISTA (*Google TAG*), Nodaria (*Symantec*), TA471 (*Proofpoint*), UAC-0056 (CERT-UA), UNC2589 (*Google Mandiant*), Lorec53 (*NSFOCUS*), Lorec Bear, Bleeding Bear (*Elastic*), and Nascent Ursa (*Palo Alto*) [4, 5].

This group belongs to the so-called 'Putin's Bears' – the state-sponsored intelligence hacking groups that are subordinate to various Russian intelligence services such as the Foreign Intelligence Service (SVR), the Federal Security Service (FSB), and Russia's Main Intelligence Directorate of the General Staff (GRU).

However, the attribution wasn't so clear at the time the WhisperGate attack happened, on 13 January 2022, a month before the full-scale invasion of Russia into Ukraine. Moreover, the attackers came with 'false flags', pretending to be Polish hackers based on what they wrote on the defaced government websites in Ukrainian, Russian and Polish languages [3].

In April 2023, *Microsoft* updated the attribution from DEV-0586 to Cadet Blizzard, a newly arranged APT group associated with the Russian General Staff Main Intelligence Directorate (GRU) [6].

*CrowdStrike* [7], in its turn, stated that the WhisperGate wipers were reminiscent of VOODOO BEAR's (a.k.a. Sandworm or GRU Unit 74455 [8]) destructive NotPetya malware [9].

*ESET* joined the fray when their researchers attributed WhisperGate and the rest of the wiper attacks to Sandworm 'with varying degrees of confidence' [10]. In our research of the Russian wipers, we also attributed these destructive attacks against the critical infrastructure of Ukraine in 2022 to Sandworm, based on the similarity of tactics and techniques used. However, attribution was not a primary goal of our research, but rather an analysis of defence evasion techniques used by the Russian intelligence groups [11, 12].

So, why do we have so many different opinions on WhisperGate attribution, even though all agree on linking WhisperGate to the Russian military intelligence service (GRU)?

The problem with the attribution of state-sponsored groups lies in the fact that they usually do not leave their signatures or any other identifiers, which is, for example, common for ransomware groups, but rather attempt to embed so-called 'false flags' to cover their tracks. This means we cannot use deterministic methods to identify a threat actor based on a single attack indicator. Instead, we should consider a set of indicators and use probabilistic methods to evaluate the match of attack traces against the profiles of known APT groups. For that purpose, we can employ traditional ML classifiers, smart pattern-matching methods, and GenAI capabilities.

**WHO IS EMBER BEAR?**

Ember Bear (a.k.a. UAC-0056, Lorec53, Lorec Bear, Bleeding Bear, Saint Bear) is an adversary group that has operated against government and military organizations in Eastern Europe since early 2021, likely to collect intelligence from target networks. Ember Bear appears primarily motivated to weaponize the access and data obtained during their intrusions to support information operations (IO) aimed at creating public mistrust in targeted institutions and degrading the government's ability to counter Russian cyber operations [4, 13].

This information is provided by *MITRE*, but the most prominent attack Ember Bear is known for is WhisperGate, and many security companies started tracking Ember Bear since 13 January 2022, under different IDs.

*Microsoft* initially used the 'DEV-0586' threat actor identifier when tracking disruptive WhisperGate attacks occurring against multiple government agencies in Ukraine in mid-January 2022 [14]. Later, they renamed this Russian GRU group to Cadet Blizzard [15].

On 30 March 2022, Adam Meyers, *CrowdStrike* Senior Vice President of Intelligence, testified in front of the CHS (House Committee on Homeland Security) on Russian cyber threats to critical infrastructure. Within his testimony, Adam spoke publicly for the first time about a Russia-nexus state-sponsored actor that *CrowdStrike Intelligence* tracks as Ember Bear [16].

With the new security advisories and indictment recently published, Ember Bear is clearly attributed to GRU's Unit 29155, which is more famous for its assassination and sabotage operations abroad than for cyber attacks [17].

Here is what *Belingcat* wrote about GRU Unit 29155 [17]:

'*We have previously identified and described operations of an elite sabotage unit within GRU Unit 29155. This unit conducts clandestine operations overseas, and we have previously identified its involvement, in addition to the series of poisonings in Bulgaria, in the annexation of Crimea (2014), destabilization attempts in Moldova (2014), a failed coup in Montenegro (2016), WADA-linked surveillance operations in Switzerland (2016-2017), possible destabilization operations in Spain during the Catalonia independence referendum (2017) and the assassination attempt on former Russian spy Sergei Skripal in Salisbury, UK (2018).*'

Moreover, one of the main technical actors involved in the WhisperGate campaign was a 22-year-old Russian civilian, Amin Stigal from Grozny, Chechnya [18]. The United States Department of State is offering a reward of up to $10 million for information leading to the location of Amin Stigal [19]. Later, in 2025, *The Insider* revealed that it was actually Stigal's father, Timur, who was responsible for executing the GRU's campaigns [20].

On 31 May 2025, after receiving access to the unprotected GRU's server logs, *The Insider* revealed even more information about the team composition and their activities in cyberspace. It turns out that the cyber unit was established in 2012 and has actively been using false flags in its disinformation campaigns.

To sum up, there are at least two interesting facts associated with a rather new GRU threat actor known as Ember Bear:

- Cyber attacks are the new direction of GRU Unit 29155, known to work mostly with assassination, sabotage, and disinformation operations.
- The unit has been recruiting young hackers since 2021, but the GRU coordinators had no IT experience, which led to the data leakage.

**APT ATTRIBUTION CHALLENGE**

Attribution in the realm of advanced persistent threats (APTs) remains one of the most difficult and controversial tasks in cybersecurity research. APTs, typically linked to state-sponsored or state-tolerated actors, operate over extended periods with defined objectives such as espionage, sabotage, financial theft and information warfare. These operations are sophisticated, often stealthy, and highly contextual. Consequently, identifying the responsible actor involves not only technical investigation but also geopolitical interpretation.

One of the fundamental problems in APT attribution is the lack of standardization in how threat actors are named and categorized across the cybersecurity industry. Prominent cybersecurity vendors such as *CrowdStrike*, *Mandiant* and *Secureworks* often assign different names to the same actor or campaign. For instance, the Russian-linked group known as 'Fancy Bear' by *CrowdStrike* is referred to as 'APT28' by *Mandiant* and 'Sofacy' by *Kaspersky* [21, 22]. This naming inconsistency complicates intelligence sharing, hampers unified incident response, and creates ambiguity for both analysts and policymakers.

Adding further complexity is the fact that attribution is inherently political. Different nation states have different motivations and doctrines that shape the nature of their cyber operations. Russia, for example, is widely recognized for employing cyber capabilities as a component of hybrid warfare. Russian APTs are often aggressive, with objectives ranging from the destabilization of political systems to attacks on critical infrastructure, serving both military and political ends [23].

In contrast, China's APT landscape is dominated by cyber-espionage campaigns aimed at acquiring economic, technological, and military advantage. Groups such as APT1 and APT10 have been linked to widespread data exfiltration operations targeting intellectual property and classified research [24].

North Korea's cyber operations present a markedly different motivation model. Isolated from the global financial system and under heavy sanctions, North Korea has leveraged cyber capabilities for direct monetary gain. APT groups like Lazarus have executed sophisticated heists targeting banks, cryptocurrency exchanges and financial platforms, often blending espionage and criminal activity [25].

A further complication in attribution arises from the reuse and sharing of tools, infrastructure and techniques across different APT groups, sometimes even across nations. It is not uncommon for malware frameworks such as PlugX, Cobalt Strike, or Mimikatz to appear in operations linked to multiple actors. Whether due to deliberate deception (false-flag operations), open-source availability, or covert collaboration, this reuse blurs the lines between threat groups and can result in misattribution or underestimation of campaign scale [26].

**METHODS OF ATTRIBUTION**

Attributing cyber attacks to specific threat actors is a complex process that involves technical analysis, intelligence gathering, and investigative techniques. Here are the most common methods of attribution, along with real-world examples for each:

1. Indicators of compromise (IoCs) include file hashes, IP addresses, domain names, or malware signatures left behind in a network or system after a cyber attack. For example, the attack on *Sony Pictures* (2014) was attributed to the Lazarus Group (North Korea) after analysts identified malware, domain names and IP addresses that matched those seen in prior North Korean cyber campaigns [27].

   Also, Sandworm, attributed to the Russian intelligence, reused IP addresses and domain names across multiple attacks, including the BlackEnergy attack in 2015 against the Ukrainian power grid, linking them to the same group [28].

2. Tactics, techniques and procedures (TTPs) describe how attackers operate, including the malware and tools they use. For example, APT28 (Fancy Bear), linked to Russian military intelligence (GRU), consistently uses the Sofacy malware, spear-phishing campaigns, credential harvesting, and brute force attacks in their operations. These TTPs were seen in April 2016 in the attack against the Democratic Congressional Campaign Committee (DCCC) and the Democratic National Committee (DNC), where the APT28 group specifically used credentials stolen through a spear-phishing email to log into the DCCC network [29].

3. Malware evolution and code similarities. Malware used in different attacks often contains unique code segments that can be linked to known samples. For example, APT29 (the foreign intelligence service of the Russian Federation, a.k.a. 'SVR') uses its custom backdoor written in Go lang, called WellMess. Ember Bear group uses SaintBot backdoor. There is significant code reuse between the early Agent.BTZ, which was used in a cyber attack on the US Department of Defense in 2008, and later Snake (Uroborus) malware used by the Turla group [30].

4. Geopolitical context. The timing, target, and nature of an attack may align with the interests of a nation state, offering clues to its origin. For example, the Stuxnet attack (2010), which targeted Iran's nuclear program, was widely believed to be a joint effort by the US and Israel, aligning with their geopolitical interests to disrupt Iran's nuclear capabilities [31].

5. Language and cultural indicators, such as language markers, time zones, and cultural habits in the code or operational behaviour, can suggest an attacker's origin. For example, *Mandiant*'s APT1 report linked cyber attacks to China's People's Liberation Army (PLA) by identifying Chinese language markers in the malware code and observing operational patterns consistent with Chinese military routines [32].

6. Human intelligence (HUMINT). Information from insiders, informants or defectors can provide direct attribution. For example, in the Silk Road takedown, authorities used insider information in addition to technical means to track down Ross Ulbricht (Dread Pirate Roberts), ultimately leading to his arrest [33].

7. Open-source intelligence (OSINT). Publicly available data, including information from social media and online forums, can offer clues for attribution. In the DNC hack (2016), researchers used OSINT to link the 'Guccifer 2.0' persona (the fake Romanian hacker) to the Russian intelligence (Fancy Bear) by analysing metadata in leaked documents, which revealed Russian language and time zone clues [34].

8. Mistakes or slip-ups by attackers. Attackers sometimes make operational mistakes, exposing their identity. In the Olympic Destroyer attack (2018) against the Winter Olympics infrastructure in Pyeongchang, South Korea, attackers from the Russian GRU (Sandworm) attempted to plant false flags to mislead investigators, but mistakes, such as infrastructure reuse, led to their identification [35].

By using these methods, often in combination, cybersecurity analysts and intelligence agencies can build a strong case for attributing cyber attacks to specific actors. However, attribution can remain challenging and sometimes speculative without substantial supporting evidence.

**EMBER BEAR SPECIFIC TRAITS: HUMAN ANALYSIS**

According to the CISA report [36], similarities in Ingress Tool Transfer (T1105) and Web Service (T1102) techniques have been found. Specifically, in both cases of attacks, WhisperGate and Ember Bear, additional malware or tools were downloaded through *Discord*'s CDN. Moreover, in one of WhisperGate's URL clusters, a SaintBot backdoor was found, known to be used by both Saint Bear and Ember Bear [4].

WhisperGate operation on 13 January 2022:

- Cluster 1:
    a. `hxxps://cdn.discordapp[.]com/attachments/928503440139771947/9301086376811847 68/Tbopbh.jpg` (a resource, e.g. payload, for `stage2.exe`)

- b. `saint.exe` (a downloader, SaintBot, as detailed by CERT-UA)
- c. `puttyjejfrwu.exe`

• Cluster 2:
- a. `hxxps://cdn.discordapp[.]com/attachments/888408190625128461/8956339522477998 58/n.lashevychdirekcy.atom.gov.ua.zip` (means for sending malware in over 35 different Zip files via *Discord* links)
- b. Several *Microsoft Word* documents with macros that download `test01.exe` from `3237.site`. Once executed, `test01.exe` downloads `load2022.exe` from `smm2021.net`.

• Cluster 3:
- a. `hxxps://cdn.discordapp[.]com/attachments/945968593030496269/9459704461495091 30/Client.exe` (note: Unit 29155 cyber actors' use of `Client.exe` was confirmed as linked to the activity, but the file was not obtained for analysis and functionality cannot be confirmed.)
- b. `asd.exe` (likely a development version of `stage1.exe`)

The Ember Bear attack on the energy sector of Ukraine on 1 February 2022, reported by *Palo Alto Networks* and CERT-UA [37, 38]:

- `cdn.discordapp[.]com/attachments/853604584806285335/854020189522755604/1406.exe`
- `cdn.discordapp[.]com/attachments/908281957039869965/908282786216017990/AdobeAcrobatUpdate.msi`
- `cdn.discordapp[.]com/attachments/908281957039869965/908310733488525382/AdobeAcrobatUpdate.exe`
- `cdn.discordapp[.]com/attachments/908281957039869965/911202801416282172/AdobeAcrobatReaderUpdate.exe`
- `cdn.discordapp[.]com/attachments/908281957039869965/911383724971683862/21279102.exe`
- `cdn.discordapp[.]com/attachments/932413459872747544/932976938195238952/loader.exe`
- `cdn.discordapp[.]com/attachments/932413459872747544/938291977735266344/putty.exe`

## ATTRIBUTION WITH TRADITIONAL MACHINE LEARNING

In this section, we present the results of attribution analysis using traditional supervised ML methods such as classifiers.

The training dataset [39] included the IoCs, tactics and techniques IDs according to the *MITRE ATT&CK* Enterprise Matrix convention for the following Russian state-sponsored APT groups:

- GhostWriter
- APT28
- APT29
- Gamaredon
- InvisiMole
- Sandworm
- DragonFly
- Turla
- Wizard Spider
- Ember Bear

The ML classifiers used in the analysis were:

- KNN
- Decision Tree
- Random Forest
- AdaBoost (SAMME)
- Linear SVM
- RBF SVM
- GaussianNB
- Neural Net

The results of the classification for the WhsiperGate campaign represented by its IoCs, tactics and techniques (see Table 1 and Figure 1) show that the majority of the classifiers attributed the attack to the Sandworm group. Still, Gaussian Naive Bayes attributed the attack with high probability to Ember Bear.

```
              KNN  Decision Tree  Random Forest  AdaBoost  Linear SVM   RBF SVM  GaussianNB  Neural Net
APT28        4.081633       2.040816       2.040816  2.040816    4.081633  0.000000    0.000000    0.000000
APT29        8.163265       2.040816       2.040816  0.000000    0.000000  4.081633    0.000000    0.000000
DragonFly    8.163265       2.040816       2.040816  4.081633    2.040816  6.122449    8.163265    4.081633
EmberBear   34.693878      36.734694      30.612245  0.000000    0.000000  6.122449   91.836735    0.000000
Gamaredon    0.000000       4.081633       0.000000  0.000000    0.000000  0.000000    0.000000    0.000000
GhostWriter  0.000000       0.000000       0.000000  0.000000    0.000000  0.000000    0.000000    0.000000
InvisiMole   2.040816       2.040816       0.000000  0.000000    0.000000  0.000000    0.000000    0.000000
Sandworm    22.448980      30.612245      38.775510 93.877551   89.795918 69.387755    0.000000   91.836735
Turla        6.122449       4.081633       6.122449  0.000000    4.081633  2.040816    0.000000    4.081633
WizardSpider 14.285714      16.326531      18.367347  0.000000    0.000000 12.244898    0.000000    0.000000
```

*Table 1: Attribution results.*

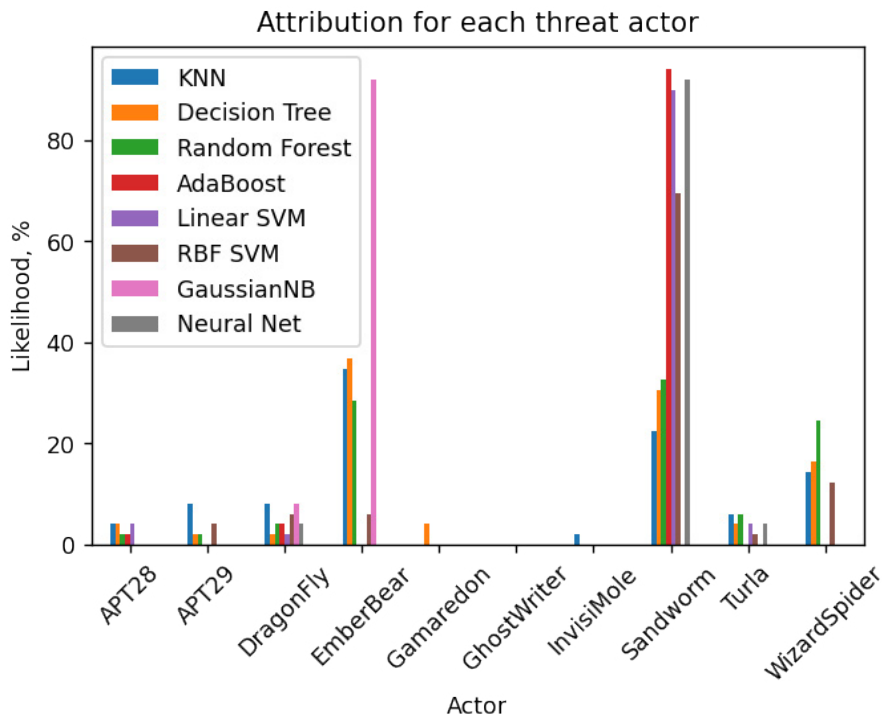

*Figure 1: The results of WhisperGate operation attribution using ML classifiers.*

## ATTRIBUTION USING CHATGPT

The WhisperGate IoCs are given to *ChatGPT 4o* within the following prompt:

> 'The file 'Unknown.csv' contains a list of indicators of compromise (IoCs) as well as tactics and techniques according to the MITRE ATT&CK enterprise matrix. Make an attribution and suggest the threat actor who is behind this attack. Provide likelihood in % for the five most probable threat actors.'

Note: the test file contains no references to any APT group.

The response:

> Based on the IoCs (Indicators of Compromise) and MITRE ATT&CK techniques you provided, here's an analysis to attribute the most likely threat actor and provide the likelihood for the five most probable threat actors.
>
> ### Key IoCs:
> - **SaintBot, WhisperGate, WhisperKill**: Known malware strains; WhisperGate and WhisperKill are associated with destructive cyberattacks, including the WhisperGate malware attack linked to Russian APTs.
> - **AdvancedRun, InstallUtil, MinGW**: Software and utilities commonly exploited by attackers for persistence and privilege escalation.

- **Programming Languages (PowerShell, VBS, C#)**: Suggest scripting capabilities, typically seen in sophisticated APT-level operations.
- **cmd.exe commands**: Several command-line indicators involving ping and network communication, process manipulation, and service stopping, hinting at lateral movement and defense evasion.
- **MITRE ATT&CK Techniques**: Techniques such as **T1134.002** (Access Token Manipulation), **T1059.001** (PowerShell), and **T1561.001** (Disk Wipe), indicate destructive activities and execution of malware scripts.

### Attribution and Threat Actors:

1. **Sandworm (APT28/Fancy Bear)**:
   - **WhisperGate**, a destructive malware similar to prior Russian wiper malware, has been tied to **Sandworm**, a unit within Russia's GRU.
   - Use of **PowerShell**, **VBS**, and **cmd.exe** aligns with Sandworm's known tactics.
   - Likelihood: **80%**

2. **APT29 (Cozy Bear)**:
   - While less associated with destructive attacks, **APT29** uses sophisticated malware, stealth, and defense evasion techniques similar to those described here (e.g., **T1059**, **T1134**).
   - Known for using **.NET** malware and command-line obfuscation.
   - Likelihood: **70%**

3. **Gamaredon**:
   - Another Russian APT group, **Gamaredon** focuses on espionage and uses wiper malware in campaigns, particularly targeting Ukrainian infrastructure.
   - Use of **InstallUtil.exe**, **MinGW**, and **PowerShell** are commonly associated with their attacks.
   - Likelihood: **65%**

4. **Wizard Spider**:
   - Involved in financially motivated and destructive attacks (e.g., **T1105**, **T1485**).
   - Known for using **PowerShell**, **cmd.exe**, and a range of malware tools.
   - Likelihood: **60%**

5. **Lazarus Group**:
   - While North Korean, **Lazarus** is known for destructive malware and espionage, including the use of wipers like **WhisperKill** and **T1059** techniques.
   - Involves command-line tools for evasion and lateral movement.
   - Likelihood: **55%**

### Conclusion:

The attack has the strongest indicators of being conducted by **Sandworm (APT28/Fancy Bear)**, based on the use of destructive malware like **WhisperGate** and techniques related to Russian wiper attacks. **APT29** and **Gamaredon** are also likely involved based on similarities in the use of malware and tactics.

This multi-faceted attack seems to align closely with Russian state-sponsored APT groups known for espionage and destructive cyber operations. If you need more detailed analysis or additional insights, feel free to ask!"

After that, *ChatGPT* was instructed to visualize the results as a bar diagram:

'Draw the bar diagram where the horizontal axis represents APT group names, and the vertical axis - likelihood.'

*ChatGPT*'s reply:

*# Re-attempt to generate the bar plot*

*plt.figure(figsize=(10, 6))*

*plt.bar(apt_groups, likelihoods, color='lightblue')*

*plt.xlabel('APT Group Names')*

*plt.ylabel('Likelihood (%)')*

*plt.title('Likelihood of Threat Actor Attribution')*

*plt.xticks(rotation=45, ha="right")*

*plt.tight_layout()*

*# Show the plot*

*plt.show()*

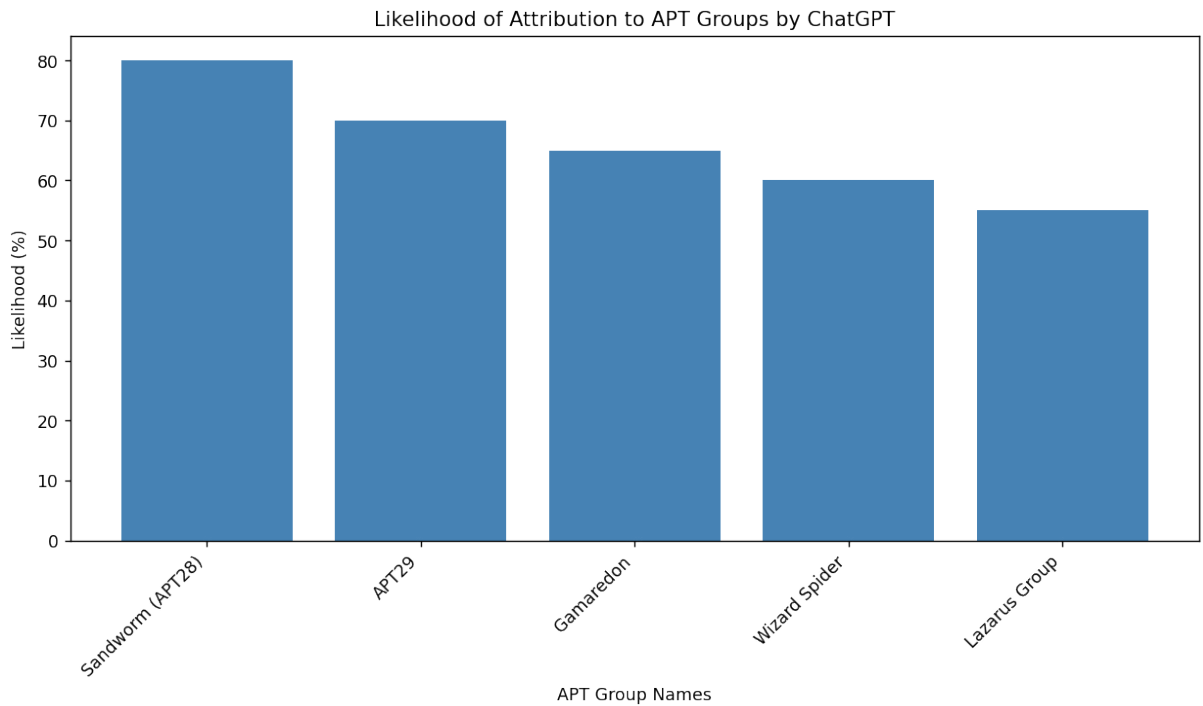

*Figure 2: The results of WhisperGate operation attribution using ChatGPT 4o.*

**FINE-TUNING**

Fine-tuning large language models can significantly improve their performance by adapting them to specific tasks, domains, or user preferences. There are many fine-tuning approaches, but the most basic is a simple (pre-)prompt engineering, using which you can instruct the LLM how to answer your questions and what data to take into account. Simply bringing information about Ember Bear operations to the current LLM's context can impact the accuracy.

Another popular approach that can be easily adopted is retrieval-augmented generation (RAG). RAG is a hybrid approach that combines retrieval-based methods and generative models to improve the performance of language models, especially in tasks that require up-to-date or factual information. It enhances the capabilities of large language models by grounding their responses in external data sources. RAG addresses this by introducing an external retrieval mechanism. When given a query, the model:

1. Retrieves relevant documents from an external knowledge base (e.g. a database, set of files, or search index).
2. Augments the model input with the retrieved documents.
3. Generates a response that incorporates the retrieved information.

The following is the answer from *ChatGPT 4o* after introducing the Ember Bear intelligence reports to the current context:

*Given the additional information that the data is associated with **Ember Bear (SaintBear)**, I will include **Ember Bear** in the list of potential threat actors. Here's the revised analysis with likelihood percentages for the five most probable threat actors behind this attack:*

*### Key IoCs and Techniques:*

  - *\*\*Malware\*\*: SaintBot, GraphSteel, GrimPlant, WhisperGate, OutSteel*

  - *\*\*MITRE ATT&CK Techniques\*\*: T1059.001 (PowerShell), T1059.003 (Windows Command Shell), T1105 (Ingress Tool Transfer), T1566.001 (Phishing), T1204.002 (Malicious File), T1110.003 (Password Spraying)*

  - *\*\*IP addresses and domains\*\*: Various associated with known Russian military cyber operations*

- **Software**: Acunetix, Amass, Metasploit, LinPEAS, Nmap

### Likelihood Attribution:

1. **Ember Bear (SaintBear)**:
    - **Likelihood**: **85%**
    - Reasoning: The presence of SaintBot, GraphSteel, and GrimPlant, which are all strongly linked to **Ember Bear**, along with the phishing and PowerShell techniques, makes it highly probable that **Ember Bear** is behind this attack.

2. **Sandworm (APT28)**:
    - **Likelihood**: **75%**
    - Reasoning: Sandworm, a Russian APT associated with destructive malware and wiper attacks, shares many of the tactics and tools used by Ember Bear, especially in campaigns targeting Ukraine.

3. **APT29 (Cozy Bear)**:
    - **Likelihood**: **65%**
    - Reasoning: While **APT29** is more focused on espionage, the use of PowerShell, phishing techniques, and command-line scripting aligns with their known behavior. However, they are less likely to use destructive malware like WhisperGate.

4. **Wizard Spider**:
    - **Likelihood**: **60%**
    - Reasoning: **Wizard Spider** is involved in financially motivated and ransomware-based campaigns. Although some overlap exists in terms of tools (like Metasploit), their operations are less aligned with the espionage and destructive focus of **Ember Bear**.

5. **Gamaredon**:
    - **Likelihood**: **55%**
    - Reasoning: Another Russian APT, **Gamaredon** frequently targets Ukraine with phishing and espionage campaigns. However, their known tooling and techniques differ slightly from those used in this attack.

### Conclusion:

Based on the indicators and techniques, **Ember Bear (SaintBear)** is the most likely threat actor behind this attack, with a high probability of 85%. **Sandworm** follows closely behind, while **APT29**, **Wizard Spider**, and **Gamaredon** are also possible, but less probable.

If you'd like further analysis or details, feel free to ask!

The generated diagram:

# Data for APT groups and their likelihoods

apt_groups = ['Ember Bear', 'Sandworm (APT28)', 'APT29', 'Wizard Spider', 'Gamaredon']

likelihoods = [85, 75, 65, 60, 55]

# Create the bar plot

plt.figure(figsize=(10, 6))

plt.bar(apt_groups, likelihoods, color='violet')

plt.xlabel('APT Group Names')

plt.ylabel('Likelihood (%)')

plt.title('Likelihood of Attribution to APT Groups')

plt.xticks(rotation=45, ha="right")

plt.tight_layout()

# Display the plot

plt.show()

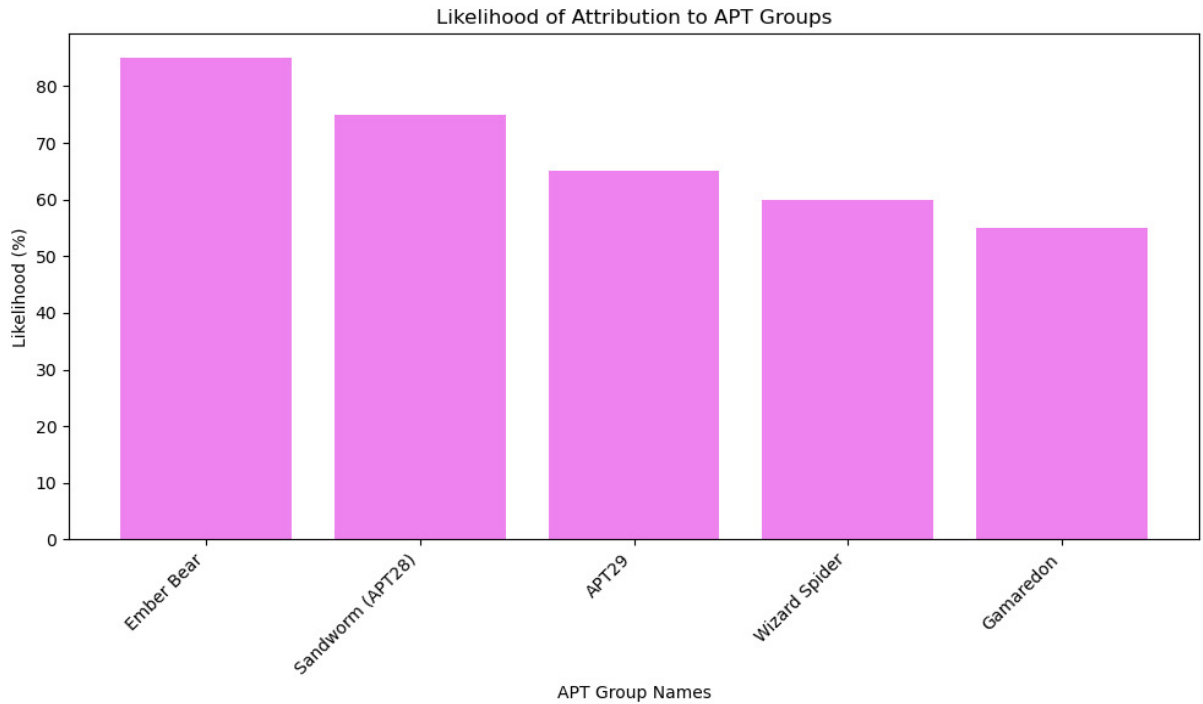

*Figure 3: The results of WhisperGate operation attribution using the fine-tuned ChatGPT 4o.*

**CONCLUSION**

This study provides a comprehensive exploration of the complexities of attribution using the WhisperGate campaign for the case study. Our investigation confirmed that traditional machine learning classifiers, as well as fine-tuned LLMs, can be effective in solving the real-world challenge of attributing APTs based on known IoCs, tactics and techniques, particularly when threat actors share the infrastructure or introduce false flags.

Both traditional ML classifiers and *ChatGPT 4o* were able, with various probabilities, to correctly attribute WhisperGate to Ember Bear without knowing the information from the published documents [1, 20] where WhisperGate operation is explicitly attributed to Ember Bear.

The integration of GenAI, accompanied by retrieval-augmented generation (RAG) and context-enriched inference, demonstrates promising enhancements affecting attribution accuracy, speed and flexibility. It brings an opportunity to incorporate other attribution methods that require semantic or linguistic analysis.

Ultimately, our work underscores the necessity of using a multifaceted attribution approach combining machine learning, geopolitical context and LLMs to navigate the uncertain landscape of state-sponsored cyber attacks.